\documentclass[superscriptaddress,amssymb,amsmath]{revtex4}

\usepackage{graphicx}
\usepackage{natbib}
\usepackage{psfrag}
\usepackage{marvosym}
\usepackage{amsmath}
\usepackage{amssymb}
\usepackage{subfigure}

\newcommand\Real{\mbox{Re}} 
\newcommand\Rey{\mbox{\textit{Re}}}  
\newcommand\upi{\pi}

\newsavebox{\astrutbox}
\sbox{\astrutbox}{\rule[-5pt]{0pt}{20pt}}

\begin{document}

\title{Self-similar vortex-induced vibrations of a hanging string}
\author{Cl\'ement~Grouthier}
\email{clement.grouthier@ladhyx.polytechnique.fr}
\affiliation{LadHyX, Department of Mechanics, Ecole Polytechnique, 91128 Palaiseau, France}
\author{S\'ebastien~Michelin}
\email{sebastien.michelin@ladhyx.polytechnique.fr}
\affiliation{LadHyX, Department of Mechanics, Ecole Polytechnique, 91128 Palaiseau, France}
\author{Yahya Modarres-Sadeghi}
\email{modarres@engin.umass.edu}
\affiliation{Department of Mechanical and Industrial Engineering, University of Massachussets, Amherst, MA01003, USA}
\author{Emmanuel~de~Langre}
\email{delangre@ladhyx.polytechnique.fr}
\affiliation{LadHyX, Department of Mechanics, Ecole Polytechnique, 91128 Palaiseau, France}
\date{\today}


\begin{abstract}
An experimental analysis of the vortex-induced vibrations of a hanging string with variable tension along its length is presented in this paper. It is shown that standing waves develop along the hanging string. The evolution of the Strouhal number $St$ with the Reynolds number $\Rey$ first follows a trend similar to what is observed for a circular cylinder in a flow for relatively low Reynolds numbers (32 $<$ $\Rey$ $<$ 700). Second, the extracted mode shapes are self-similar : a rescaling of the spanwise coordinate by a self-similarity coefficient allows all of them to collapse on a unique function. The self-similar behaviour of the spatial distribution of the vibrations along the hanging string is then explained theoretically by performing a linear stability analysis of an adapted wake-oscillator model. This linear stability analysis finally provides an accurate description of the mode shapes and of the evolution of the self-similarity coefficient with the flow speed.
\end{abstract}

\maketitle


\section{Introduction}

Vortex-induced vibrations (VIV) of slender structures are of major importance in many engineering fields. In offshore applications for instance, a good knowledge of the dynamic loads exerted on risers and mooring cables is required to prevent fatigue damage \citep{Baar,Moda2}. VIV are self-sustained oscillations of an immersed bluff body originating in a strong coupling between the structure's dynamics and its induced fluctuating wake. This coupling may lead to a synchronization of the solid vibrations and vortex shedding frequencies, a phenomenon known as lock-in, resulting in large amplitude oscillations \citep{Blev,Will}.

The case of long and flexible structures in VIV, most relevant to offshore applications, has been extensively studied using numerical computations \citep{Newm,Evan}, experiments \citep{Vand,Moda} and reduced-order models \citep{Viol,Srin}. When a long flexible structure is placed perpendicular to the oncoming flow, higher vibration modes can be excited as the flow velocity is increased, through the successive synchronizations of its structural modes with the local shedding frequency \citep{Chap}. Spatial variations of the vortex shedding frequency, for example in sheared flows, significantly modifies the dynamics from the canonical problem of tensioned cables in uniform flows \citep{Mathel,Bour,Bour2}. Similarly, spatial variations of the cable's tension (for example due to gravity and/or free end conditions) is expected to impact the structure's dynamics, in particular by localizing vibrations in regions with lower tension. This second configuration has so far received little attention \citep{Park,Srin2}. The present paper thus discusses an experimental investigation of the VIV of a hanging string, where gravity-induced tension does vary from bottom to the top. 

Section \ref{sec:Exp} presents the experimental setup and typical raw measurements. In Section \ref{sec:Strouhal}, the evolution of the dimensionless frequency of oscillations with Reynolds number is analyzed and Section \ref{sec:mode_shape} identifies a self-similar behaviour in the spatial distribution of the amplitude of vibrations. This self-similarity is confirmed analytically in Section \ref{sec:section_model} using a wake-oscillator model, before concluding remarks are given in Section \ref{sec:conclusion}.

\section{Experiments}\label{sec:Exp}

\subsection{Experimental setup}

\begin{figure}
	\psfrag{d}[cc][cc][0.75]{$d$}
	\psfrag{D}[cc][cc][0.75]{$D$}
	\psfrag{De}[cc][cc][0.75]{$D_{eq}$}	
	\psfrag{Cam}[lc][cc][0.75]{Camera}		
	\psfrag{U}[cc][cc][0.75]{$U$}
	\psfrag{L}[cc][cc][0.75]{$L$}	
	\psfrag{X}[lc][cc][0.75]{$X$}	
	\psfrag{Y}[rc][cc][0.75]{$Y$}	
	\psfrag{Z}[bc][cc][0.75]{$Z$}			
	\psfrag{a}[rc][lc][1]{$(a)$}  				
	\centerline{\includegraphics[height = 6cm]{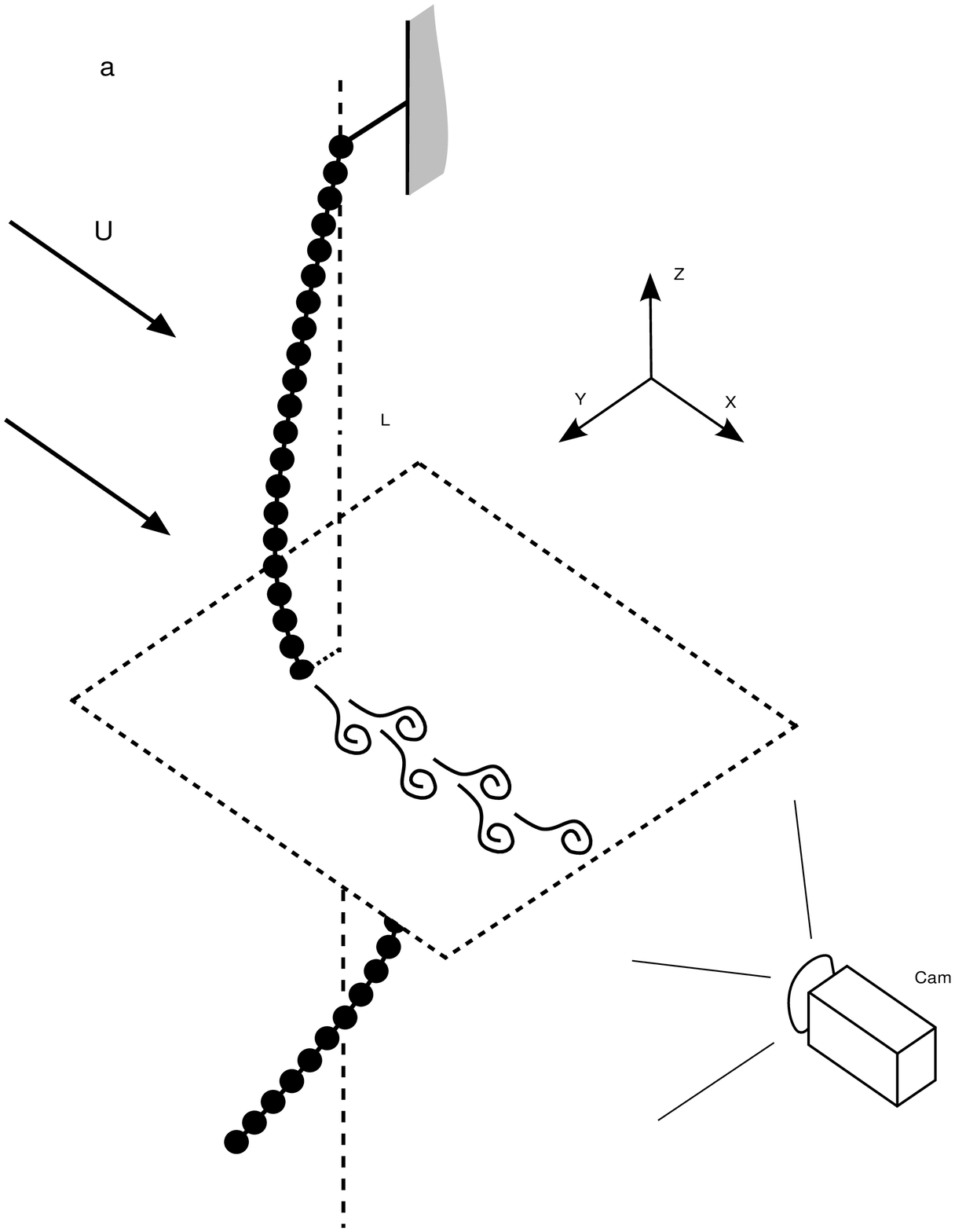} \hspace{1cm}
	\psfrag{b}[rc][lc][1]{$(b)$}  		
	\psfrag{c}[lc][rc][1]{$(c)$}  	
	\psfrag{h}[bc][cc][0.75]{$h$}  	
  \includegraphics[height = 6cm]{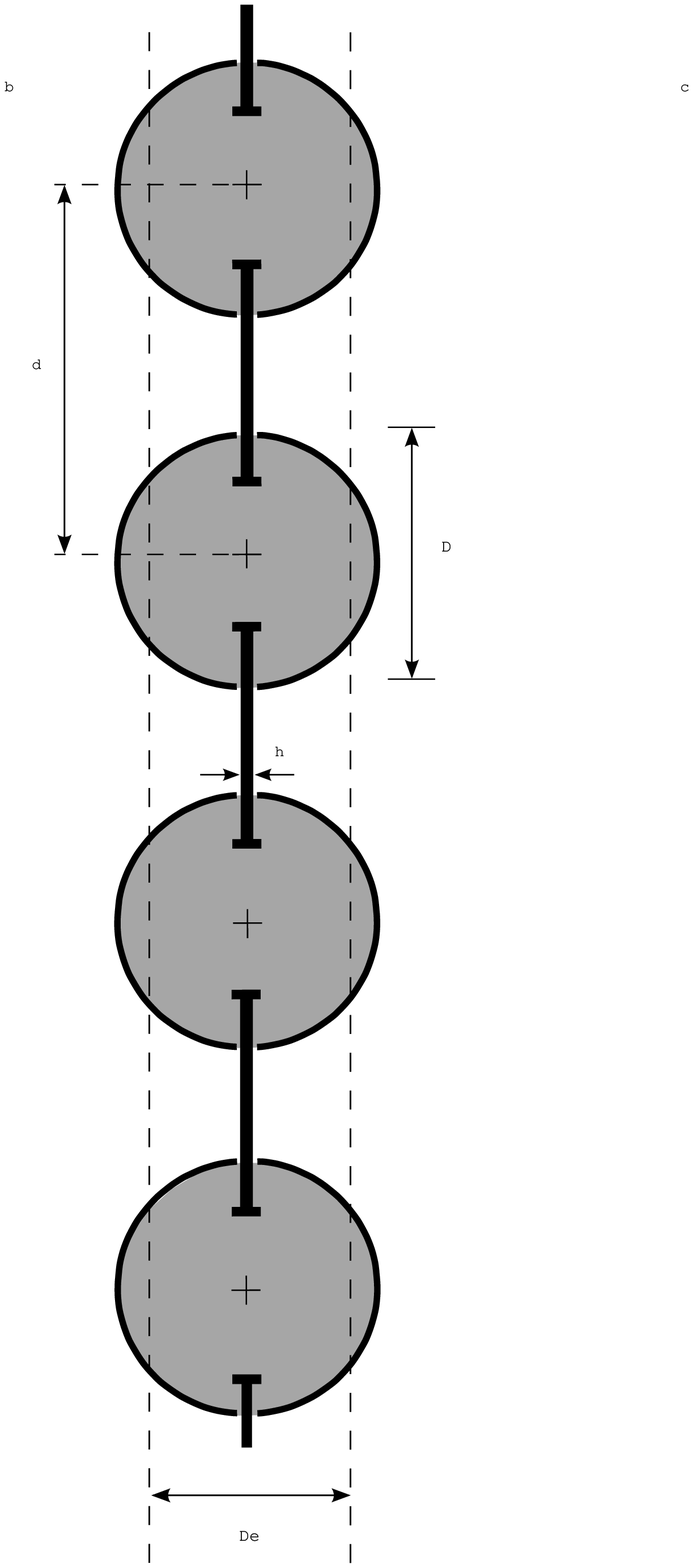} \hspace{1cm}
  \includegraphics[height = 6cm]{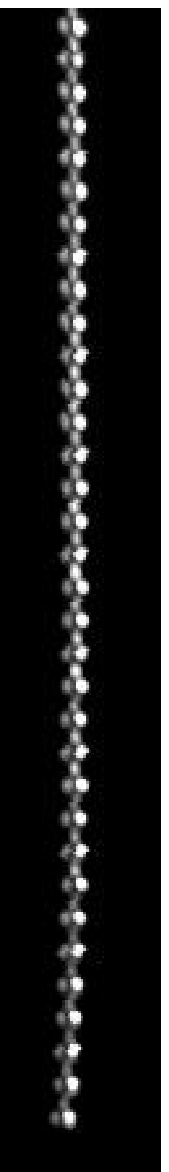}}
  \caption{$(a)$ Experimental setup of the hanging string subject to a transverse flow. $(b)$ Geometry of the chains that were used to model a string of equivalent diameter $D_{eq}$. $(c)$ Visualization of Chain 3 during one typical experiment.}
\label{schema_chainette}
\end{figure}

Experiments were performed in the Fluid-Structure Interactions laboratory of the University of Massachussets in Amherst. Fixed at its top extremity, a chain was hung in the water channel and was submitted to a flow of constant speed $U$, figure \ref{schema_chainette} $(a)$. The chain consisted of several spheres of equal diameter connected through short metallic rods, figure \ref{schema_chainette} $(b)$. Three different chains were used, and their respective characteristics are shown in table \ref{tab_chain} where $L$ and $D$ are the chain length and diameter, $d$ is the distance between two successive spheres, $h$ the diameter of the connecting metallic rods and $m_{s}$ the chain mass per unit length. The cross-flow displacement $Y \!\left( Z, T \right)$ was recorded using a Phantom Miro M110 camera through a window located at the downstream end of the water channel, figure \ref{schema_chainette} $(a)$. The camera was equipped with a lens of focal length 50 mm and aperture f/1.4. The position of the chain was obtained by tracking the contrast between the bright spheres and dark background, figure \ref{schema_chainette} $(c)$, and filtering the resulting displacement signal around its dominant frequency. Some wake visualizations were also performed using dye released slightly upstream from the lower tip of the string.

The chains used in the experiments were heavy enough not to float while keeping a very small bending stiffness for small curvatures. Yet, significant non-linear rigidity may appear for higher curvatures. When the wavelength is large in comparison with the spheres spacing $d$, the chain can be used to model an ideal string of equivalent diameter $D_{eq} = \upi D /4$. Here, $D_{eq}$ is defined such that the equivalent string has the same cross-sectional area as the spheres, thereby neglecting the effect of the connecting rods on the fluid/solid coupling. The Reynolds number $\Rey$ is based on this equivalent diameter and reads as

\begin{equation}
	\Rey = \frac{U D_{eq}}{\nu}, 
	\label{D_eq}
\end{equation}

\noindent{where $\nu = 10^{-6} \mbox{ m}^{2}\mbox{.s}^{-1}$ is the kinematic viscosity of water at room temperature . With this definition, the Reynolds number varies in these experiments between 32 and 700.}

For all experiments, the flow velocity was sufficiently low for the string's static in-line deflection to be negligible, so that the flow may be considered perpendicular to the string. The in-line vibrations were too small to be detected in our setup and are therefore not investigated further, although they are known to exist in such configurations \citep{Huera}. The present study consequently focuses on the string's cross-flow oscillations. Finally, it ought to be mentioned here that it was not possible to capture the motion of the top 6 cm (of the total 48cm) of the string because of the size of the window through which videos were recorded. This does not affect the conclusions and explains why the top displacement of the string is not shown later in figure \ref{Disp_brut} $(a)$ and $(b)$.

\begin{table}
  \begin{center}
\def~{\hphantom{0}}
  \begin{tabular}{lccccc}
      		      &   $L$ (m)&   $D$ (mm)   &   $d$ (mm)   &   $h$ (mm)  &  $m_{s}$ (g/m)  \\[3pt]
       Chain 1  &   0.48   &   2.0   &   3.0   &   0.4   &   8   \\
       Chain 2  &   0.48   &   3.1   &   4.5   &   0.6   &   21  \\
       Chain 3  &   0.48   &   4.4   &   6.3   &   0.8   &   29  \\
  \end{tabular}
  \caption{Dimensional properties of the three different chains used for experiments.}
  \label{tab_chain}
  \end{center}
\end{table}

\subsection{Wake visualizations}

\begin{figure}
	\psfrag{a}[cc][cc][1]{$(a)$}
	\psfrag{b}[cc][cc][1]{$(b)$}
	\psfrag{c}[cc][cc][1]{$(c)$}
	\psfrag{d}[cc][cc][1]{$(d)$} 		
  \centerline{\includegraphics[width = 0.75\linewidth]{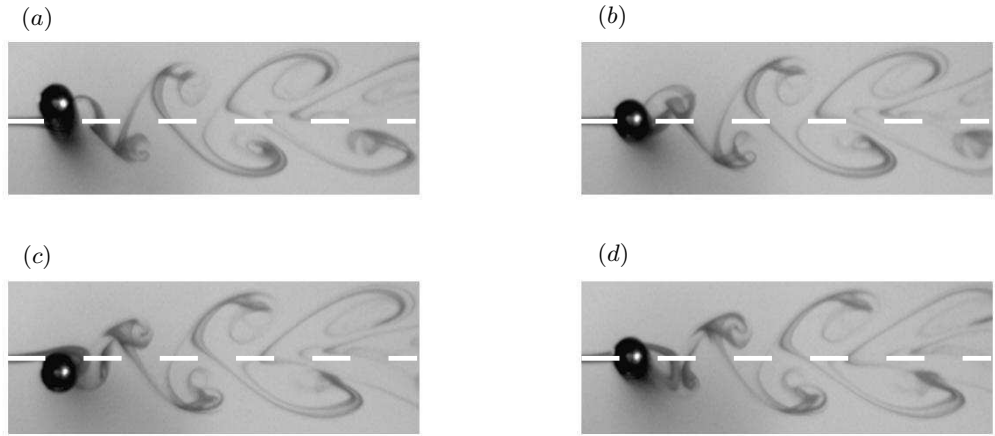}}
  \caption{Wake visualization during one cycle of oscillations of chain 2 for $\Rey = 106$. The white dashed line on each picture shows the location of the chain at rest. $(a)$ and $(c)$ correspond to the minimum and maximum displacement of the chain respectively.}
\label{Wake}
\end{figure}

Wake visualizations were performed in order to confirm that the measured displacements actually are vortex-induced vibrations. Four snapshots showing the tip displacement of the hanging string as well as its near wake are shown in figure \ref{Wake}. The vortex shedding behind the string is synchronized with the periodic oscillations of the solid: one vortex is shed every time the free end of the hanging string reaches its minimum or maximum displacement, as shown in figure \ref{Wake}. The frequency of vibrations of the body and the frequency of the vortex shedding are locked, and the vibrations result from this strong coupling between the string's dynamics and its induced wake.

\subsection{String's motion}

\begin{figure}
\centering
	\psfrag{T}[tc][cc][1]{$T$(s)}	
	\psfrag{Z}[bc][tc][1][-90]{$Z$(m)}			
	\psfrag{y0}[cc][cc][0.75]{0}		
	\psfrag{Y}[bc][tc][1][-90]{$y$}		
	\psfrag{y2}[rc][cc][0.75]{0.20}
	\psfrag{y4}[rc][cc][0.75]{0.40}
	\psfrag{yf}[rc][cc][0.75]{0.48}		
  \psfrag{c0}[cc][cc][0.75]{0}
  \begin{minipage}[c]{.45\linewidth}
		\psfrag{x0}[cc][cc][0.75]{0}
		\psfrag{x1}[cc][cc][0.75]{1}
		\psfrag{x2}[cc][cc][0.75]{2}
		\psfrag{x3}[cc][cc][0.75]{3}
		\psfrag{x4}[cc][cc][0.75]{4}
		\psfrag{x5}[cc][cc][0.75]{5}  
  	\psfrag{cm5}[lc][cc][0.75]{-0.45}   
  	\psfrag{c45}[lc][cc][0.75]{0.45}  
		\psfrag{a}[rc][lc][1]{$(a)$}  	
    \includegraphics[width = 0.95\linewidth]{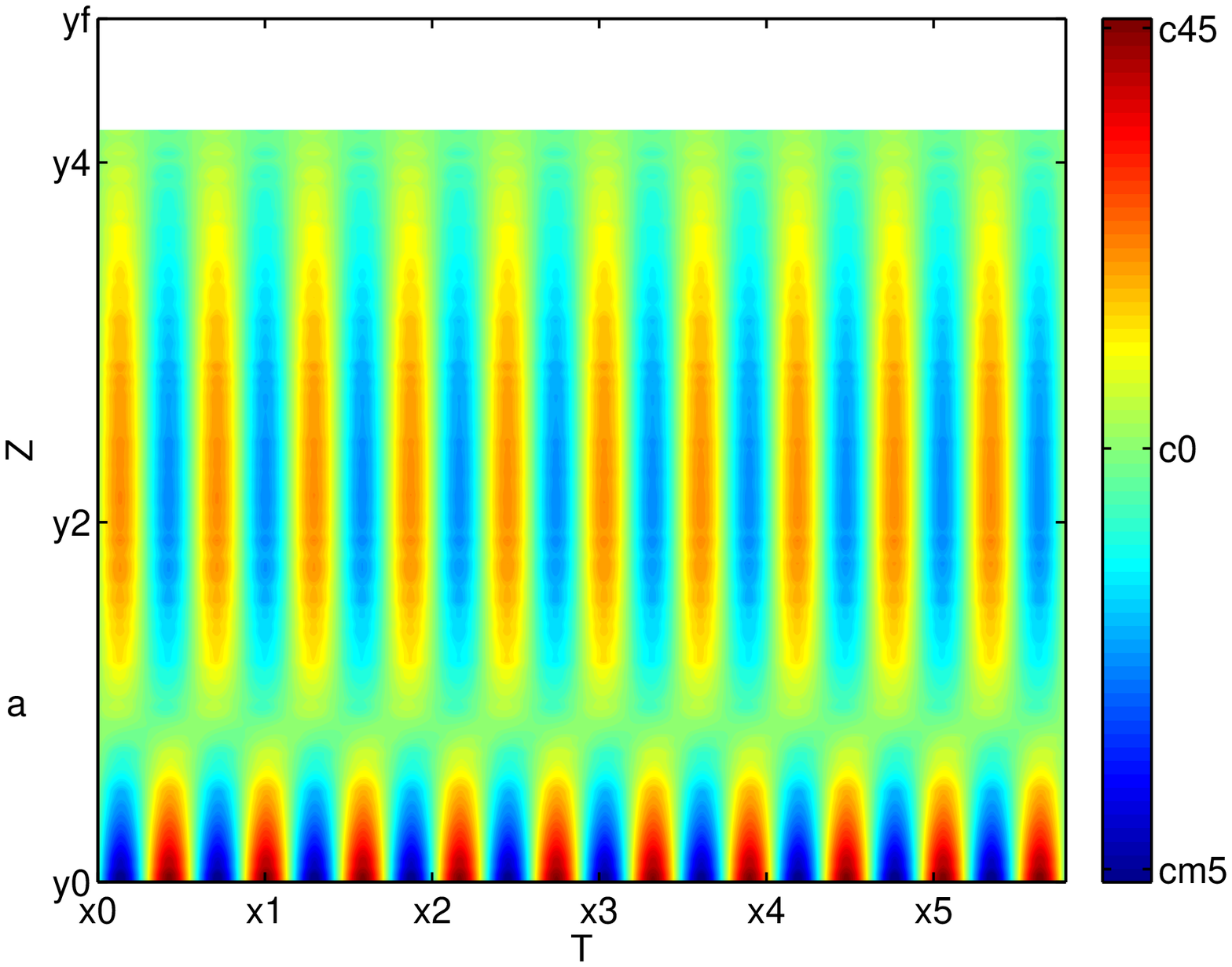}
    \vspace{0.3cm}
  \end{minipage} \hspace{0.3cm}
  \begin{minipage}[c]{.45\linewidth}
  	\psfrag{x0}[cc][cc][0.75]{0}
		\psfrag{x1}[cc][cc][0.75]{1}
		\psfrag{x2}[cc][cc][0.75]{2}
  	\psfrag{cm5}[lc][cc][0.75]{-0.55}   
  	\psfrag{c55}[lc][cc][0.75]{0.55}  
		\psfrag{b}[rc][lc][1]{$(b)$}    		
    \includegraphics[width = 0.95\linewidth]{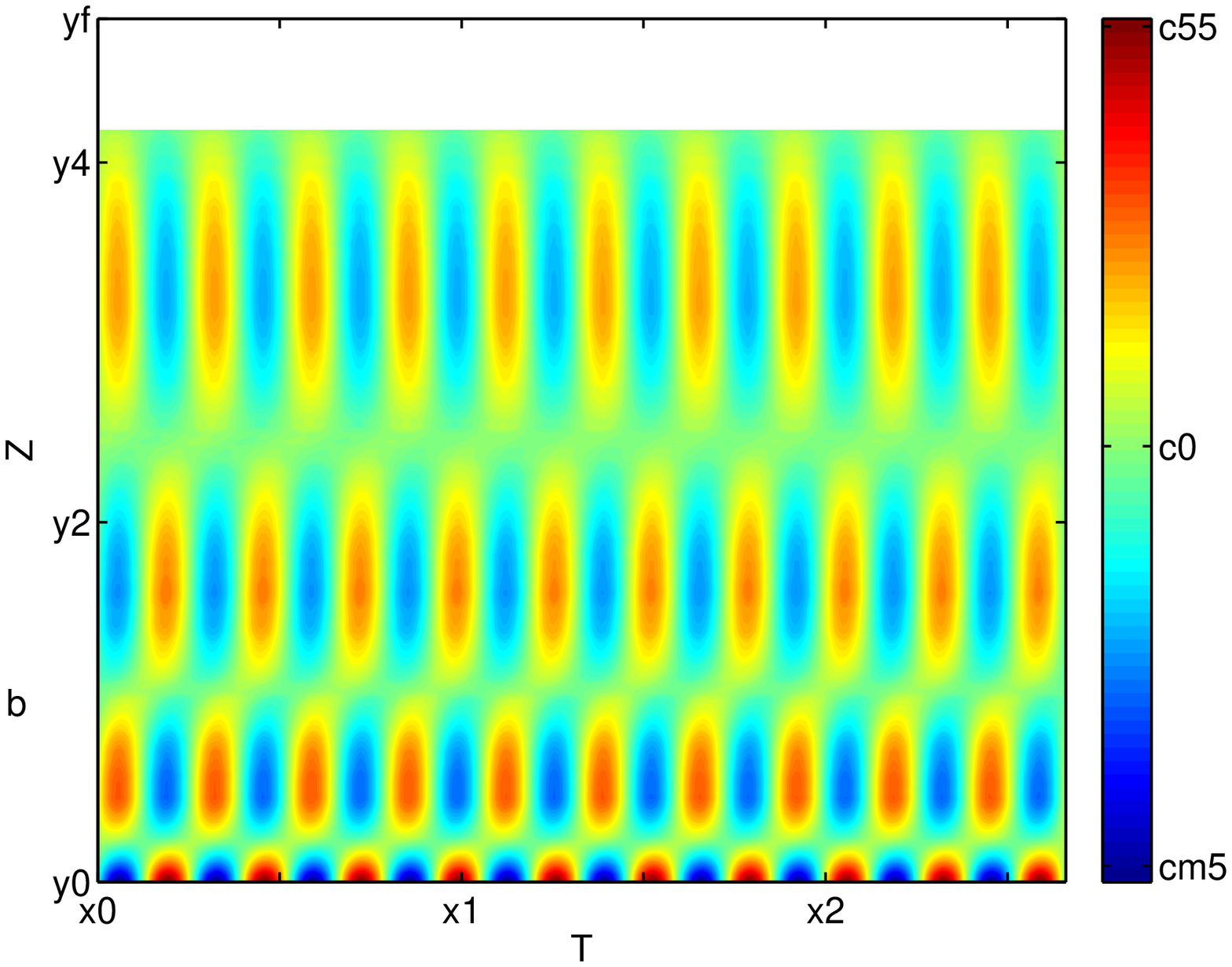}
    \vspace{0.3cm}
  \end{minipage} \vfill
  \begin{minipage}[l]{.45\linewidth}
    \hspace{-0.38cm}
		\psfrag{x0}[cc][cc][0.75]{0}
		\psfrag{x1}[cc][cc][0.75]{1}
		\psfrag{x2}[cc][cc][0.75]{2}
		\psfrag{y00}[cc][cc][0.75]{0}
		\psfrag{ym1}[cc][cc][0.75]{-1}
		\psfrag{y01}[cc][cc][0.75]{1}		
		\psfrag{x3}[cc][cc][0.75]{3}
		\psfrag{x4}[cc][cc][0.75]{4}
		\psfrag{x5}[cc][cc][0.75]{5}   
		\psfrag{c}[rc][lc][1]{$(c)$}   		   
    \includegraphics[width = 0.835\linewidth]{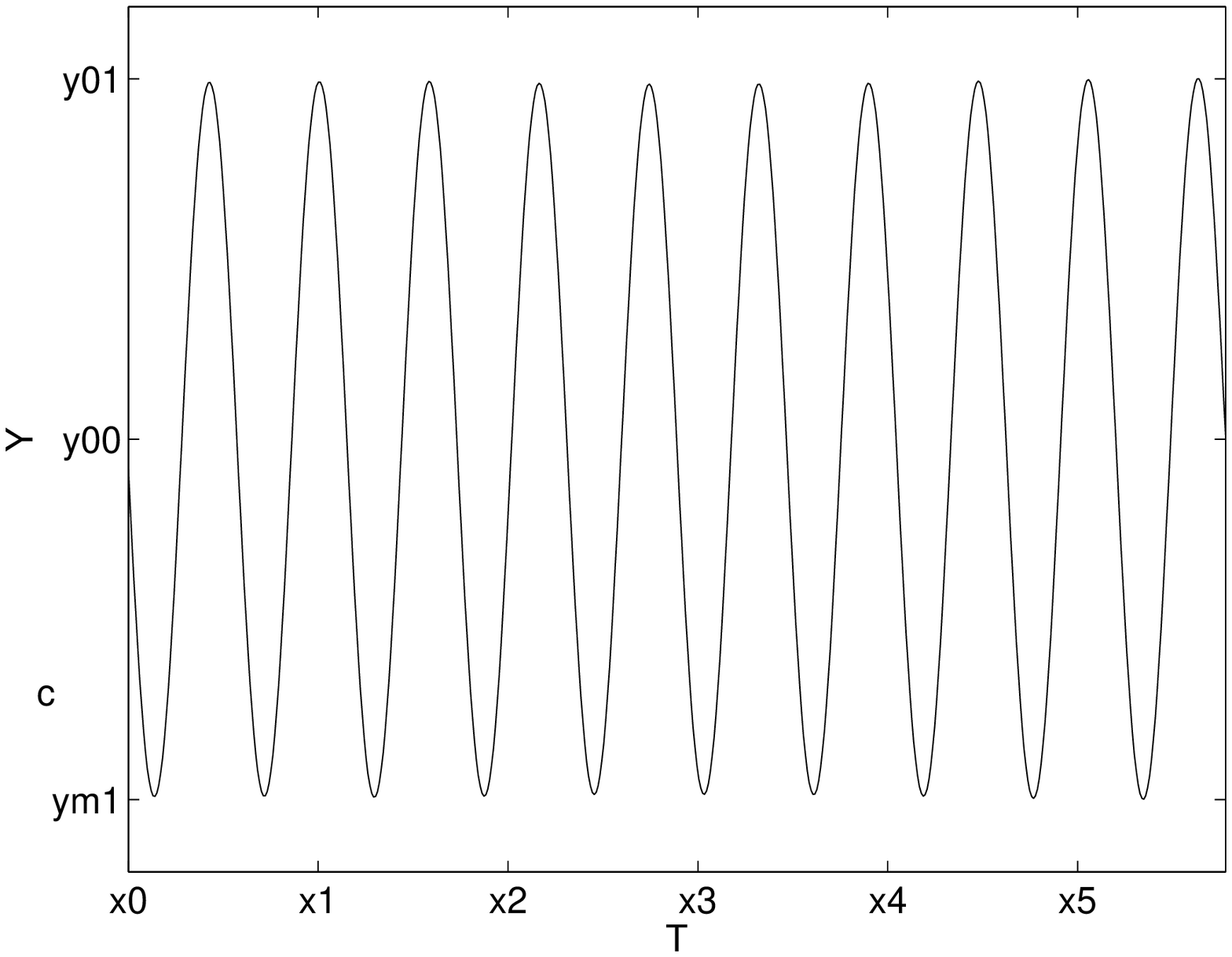}
    \vspace{0.3cm}  
  \end{minipage} 
  \begin{minipage}[c]{.45\linewidth}
  \hspace{0.025cm}
		\psfrag{x0}[cc][cc][0.75]{0}
		\psfrag{x1}[cc][cc][0.75]{1}
		\psfrag{x2}[cc][cc][0.75]{2}  
		\psfrag{d}[rc][lc][1]{$(d)$}  
		\psfrag{y00}[cc][cc][0.75]{0}
		\psfrag{ym1}[cc][cc][0.75]{-1}
		\psfrag{y01}[cc][cc][0.75]{1}				 		
    \includegraphics[width = 0.835\linewidth]{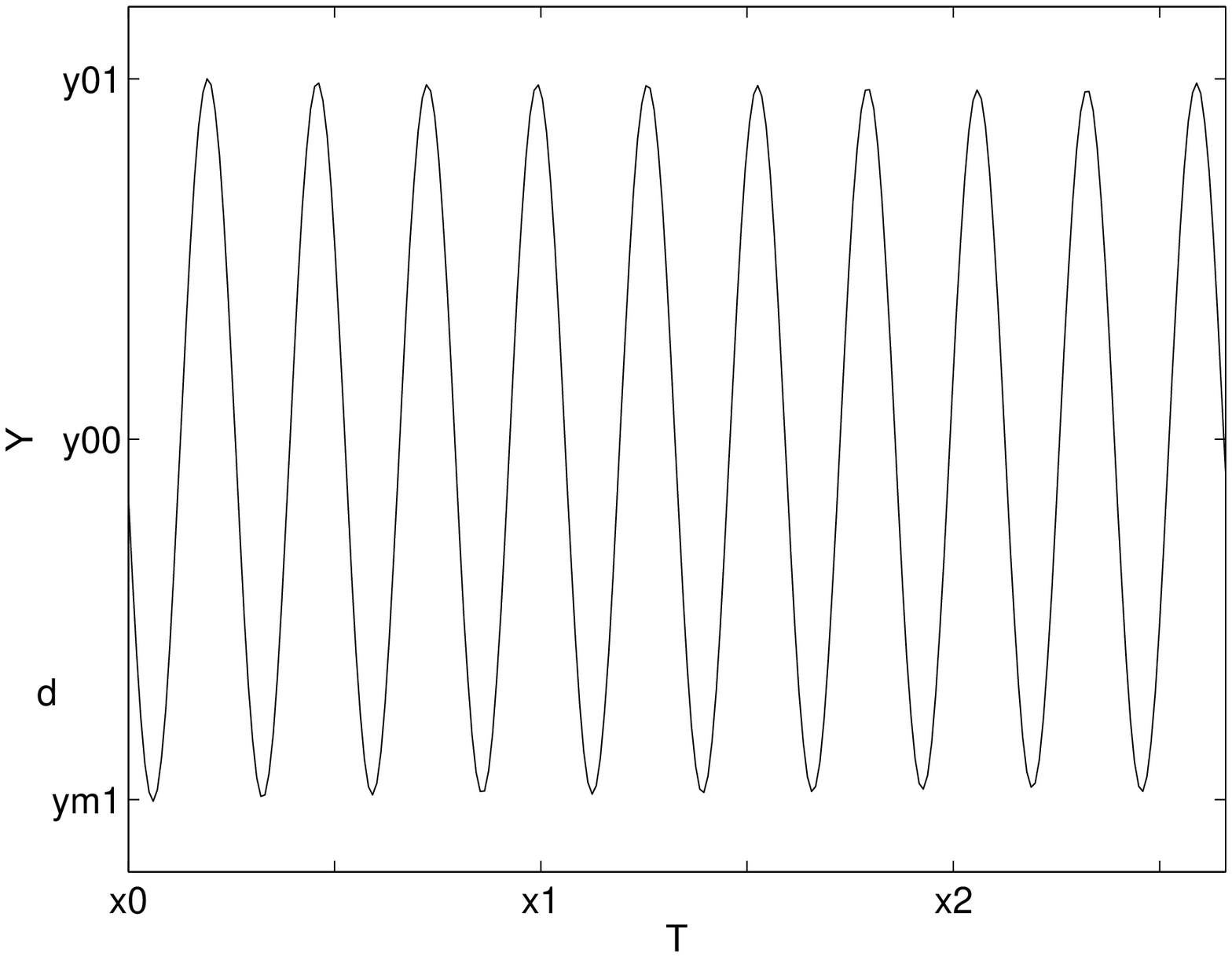}
  \vspace{0.3cm}
  \end{minipage} \vfill 
  \begin{minipage}[l]{.45\linewidth}
    \hspace{-0.25cm}
		\psfrag{x00}[cc][cc][0.75]{0}
		\psfrag{xm5}[cc][cc][0.75]{-0.5}
		\psfrag{x05}[cc][cc][0.75]{0.5}      
		\psfrag{phi}[tc][cc][1]{$y$}			
		\psfrag{e}[rc][lc][1]{$(e)$}   				      
    \includegraphics[width = 0.815\linewidth]{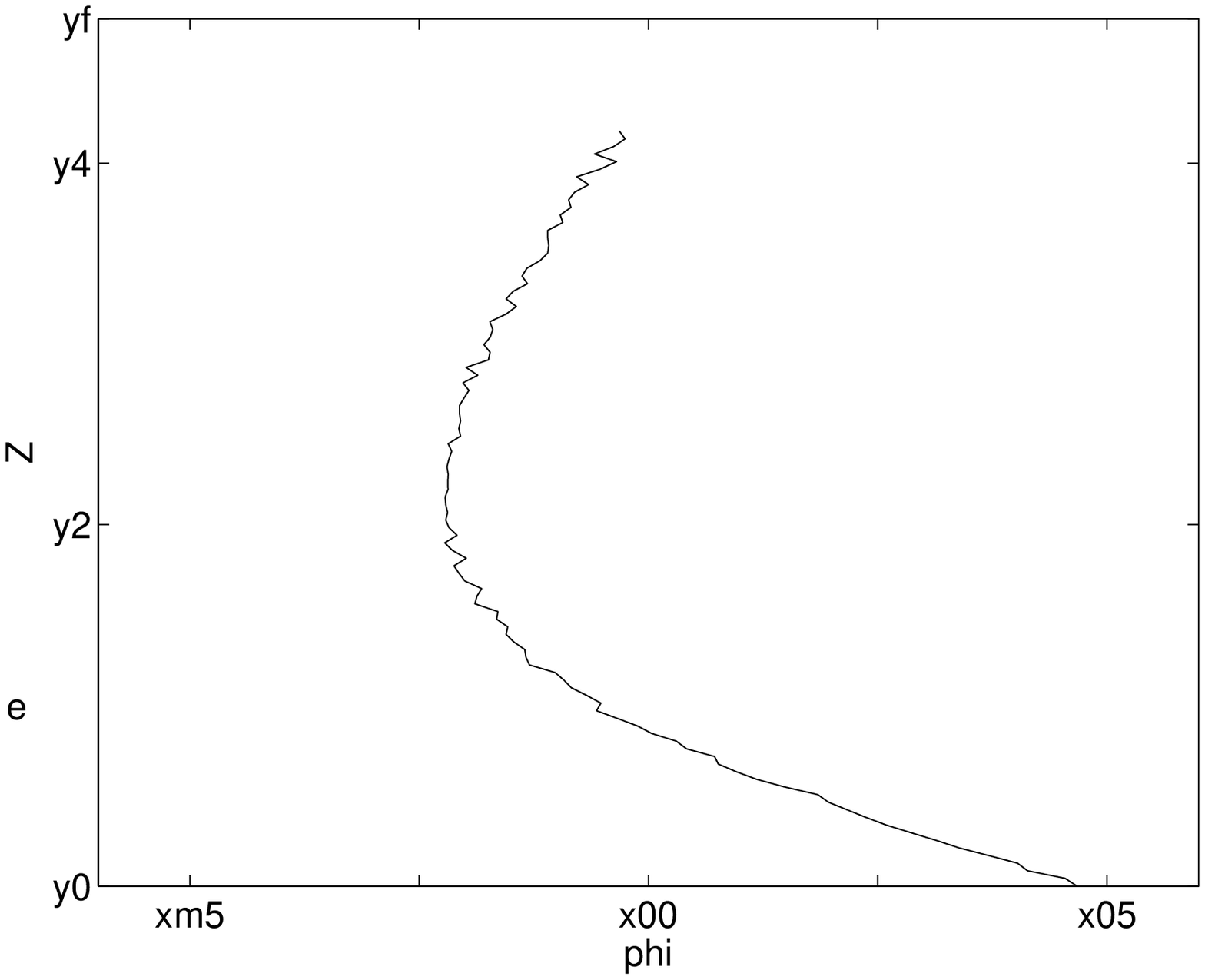}
    \vspace{0.3cm}  
  \end{minipage} 
  \begin{minipage}[c]{.45\linewidth}
  \hspace{0.155cm}
 		\psfrag{x00}[cc][cc][0.75]{0}
		\psfrag{xm5}[cc][cc][0.75]{-0.5}
		\psfrag{x05}[cc][cc][0.75]{0.5}    
		\psfrag{phi}[tc][cc][1]{$y$}				 
		\psfrag{f}[rc][lc][1]{$(f)$}   				
    \includegraphics[width = 0.815\linewidth]{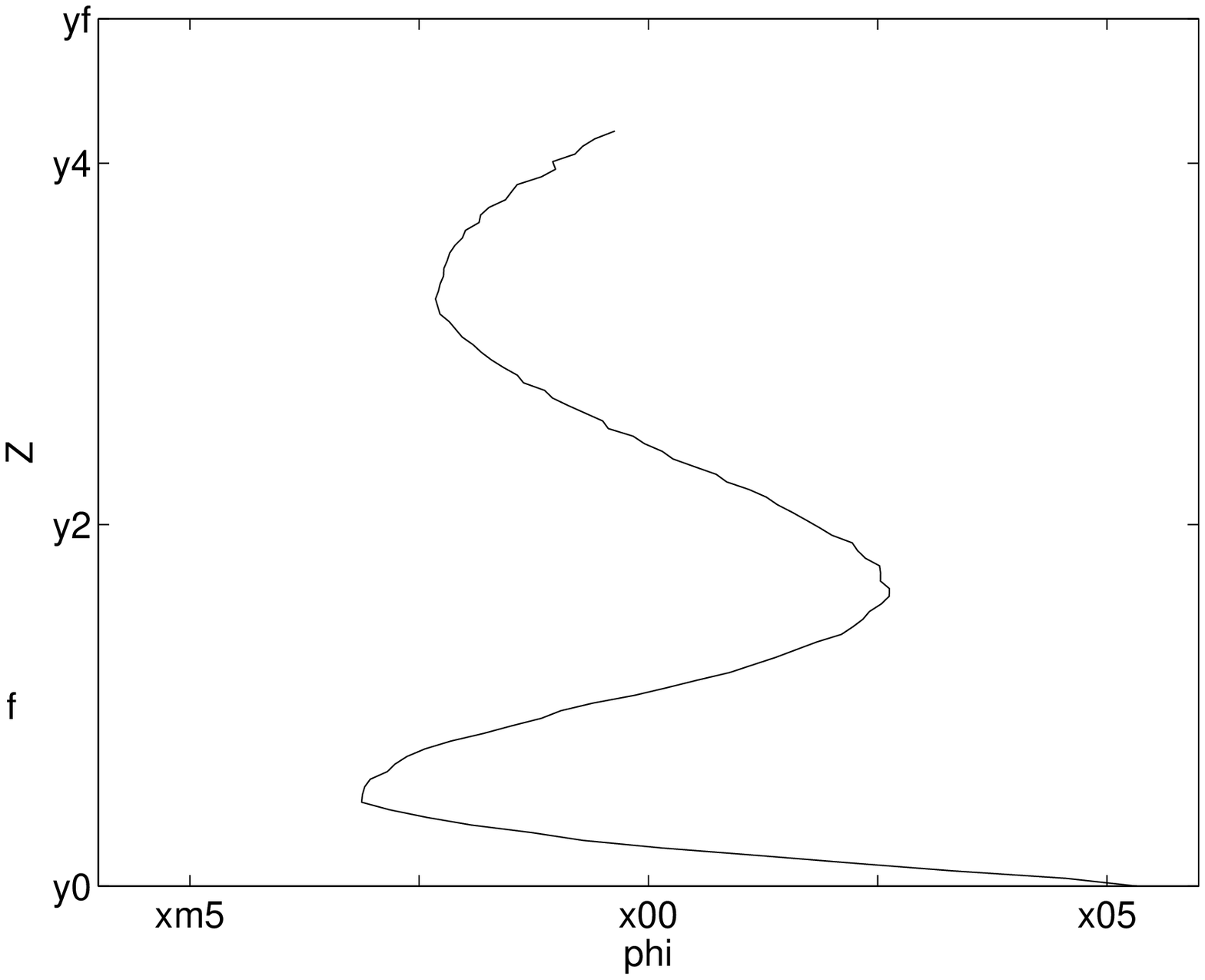}
  \vspace{0.3cm}
  \end{minipage}    
  \caption{Vortex-induced vibrations of Chain 3 for two typical experiments, $U = 0.033 \mbox{ m.s}^{-1}$ ($\Rey = 115$) (left) and $U = 0.064 \mbox{ m.s}^{-1}$ ($\Rey = 221$) (right). $(a)$,$(b)$ Dimensionless displacement $y = Y/D_{eq}$ (level step : 0.015), $(c)$,$(d)$ Rescaled tip displacement, the amplitudes of the temporal components were arbitrarily set to 1, $(e)$,$(f)$ spatial component of the solid's displacement, derived as the time-averaged shape of the string over all the time-steps for which the tip displacement is larger than 3/4 of the tip r.m.s. amplitude. }
\label{Disp_brut}
\end{figure}

The string's transverse displacement was obtained from the videos and is plotted in figures \ref{Disp_brut} $(a)$ and $(b)$ for two typical cases, showing standing waves along the hanging string. The maximum amplitudes are significant -- of the order of half of the equivalent diameter. The vibrations with the highest amplitudes are concentrated in the lower part of the string, where the tension is lower.

The standing wave can be decomposed into its temporal and spatial components, as shown in figure \ref{Disp_brut} $(c)$-$(f)$. As the flow velocity increases, a higher structural mode is excited, which corresponds to a higher frequency of oscillations, since VIV are due to the lock-in between vortex shedding and one of the string's eigenfrequencies. In the sample cases shown in figure \ref{Disp_brut}, the second and the fourth structural modes are excited, respectively, as the flow velocity increases from $U = 0.033 \mbox{ m.s}^{-1}$ to $U = 0.064 \mbox{ m.s}^{-1}$. The excitation of higher modes at higher flow velocities is expected in VIV of flexible cylinders \citep{Chap,Trim}.

\section{Strouhal number}\label{sec:Strouhal}

The natural frequency of vortex shedding, $F_{s}$, is defined here as the frequency at which vortices would be shed if the string were fixed and perfectly rigid. The Strouhal number is thus defined as

\begin{equation}
St = \frac{F_{s} D_{eq}}{U},
\label{St_def}
\end{equation}

\begin{figure}
	\psfrag{Reff}[tc][cc][1]{$\Rey$}
	\psfrag{Steff}[rc][cc][1][-90]{$St$}
	\psfrag{x0}[cc][cc][0.75]{0}
	\psfrag{x4}[cc][cc][0.75]{400}
	\psfrag{x8}[cc][cc][0.75]{800}
	\psfrag{y01}[cc][cc][0.75]{0.1}
	\psfrag{y15}[cc][cc][0.75]{0.15}	
	\psfrag{y02}[cc][cc][0.75]{0.2}
	\psfrag{y25}[cc][cc][0.75]{0.25}	
	\psfrag{203mm}[cc][cc][0.65]{$D = 2.0$ mm} 
	\psfrag{313mm}[cc][cc][0.65]{$D = 3.1$ mm}
	\psfrag{444mm}[cc][cc][0.65]{$D = 4.4$ mm}			  
	\psfrag{Fit}[lc][cc][0.65]{$0.25 - \frac{0.64}{\sqrt{\Rey}}$}		
	\psfrag{Will}[lc][cc][0.65]{$0.2698 - \frac{1.0272}{\sqrt{\Rey}}$}			
  \centerline{\includegraphics[width = 0.6\linewidth]{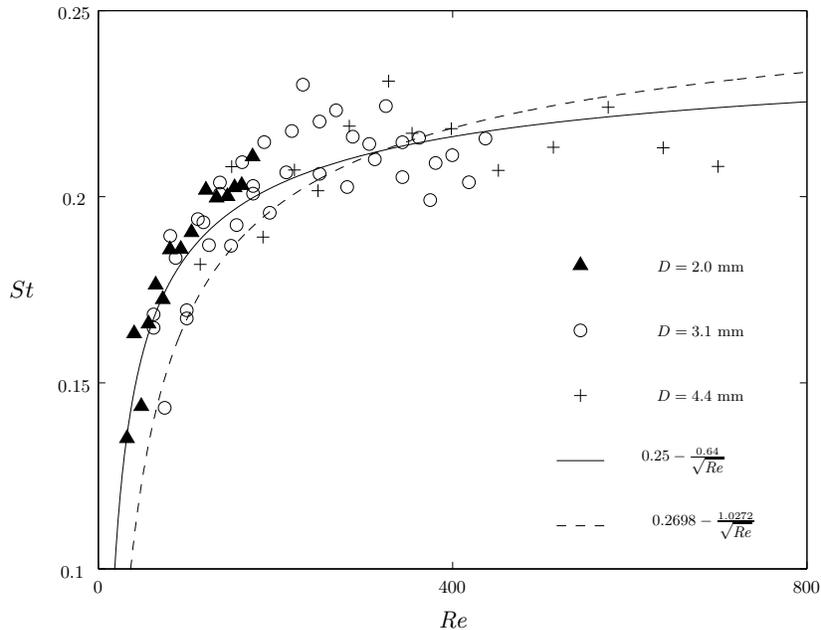}}
  \caption{Strouhal number as a function of the Reynolds number: measurements (symbols), best fit in a series in $1 / \sqrt{\Rey}$ (continuous line) and fit introduced in Ref.~\cite{Will2} (dashed line)}
\label{Strouhal_effectif}
\end{figure}

\noindent{For flexible structures, VIV originate in the lock-in between the frequency of vortex shedding and one of the eigenfrequencies of the structure \citep{Wu}. The measured frequency $F$ of the string's displacement is therefore close, but not equal to, the natural frequency of vortex shedding, $F_{s}$. The Strouhal number, as defined in equation~\eqref{St_def}, may thus be estimated here by $F D_{eq} / U$, and is shown in figure \ref{Strouhal_effectif} as a function of the Reynolds number.} 

The Strouhal number $St$ is close to 0.2, which is the value commonly used for cylinders over a wide range of $\Rey$ \citep{Blev, Naud}. Moreover, $St$ increases with $\Rey$ for $\Rey < 200$ before it reaches a constant level slightly higher than 0.2 for higher values of $\Rey$. This trend is very similar to the $St$ vs $\Rey$ plot that can be found for instance in Ref.~\cite{Blev} or in Ref.~\cite{Norb}. It was also shown in Refs.~\cite{Fey,Will2} that a fit in a series of $1 / \sqrt{\Rey}$ is appropriate to interpolate the variations of $St$ in this range of $\Rey$. Such a fit is derived for the presented data and is plotted in figure \ref{Strouhal_effectif},

\begin{equation} 
	St = 0.25 - \frac{0.64}{\sqrt{\Rey}}.
	\label{St_fit}
\end{equation}  

\noindent{The coefficients proposed in Ref.~\cite{Will2} also result in a reasonably good fit for the presented measurements, figure \ref{Strouhal_effectif}. This shows the relevance of this estimation of the Strouhal number as well as the consistency of its variations with $\Rey$. The chosen definition of $D_{eq}$ may impact the experimental curve of figure \ref{Strouhal_effectif}, therefore explaining part of its deviation from the fits derived in Ref.~\cite{Will2}.}

Equation~\eqref{St_fit} is consequently used in the remainder of the paper to estimate the Strouhal number of the discussed experimental results. 

\section{Mode shapes} \label{sec:mode_shape}

\subsection{Characteristic dimensions of the problem} \label{charac}

The equivalent diameter $D_{eq}$ is a natural characteristic length for the cross-flow displacement $Y \!\left( Z, T \right)$, and the natural frequency of vortex shedding $\omega_{s} = 2 \upi St  U / D_{eq}$ defines a characteristic frequency for the string's VIV. 

For the spanwise coordinate $Z$, a characteristic length $Z_{c}$ is derived from a typical phase speed, as follows. The tension is induced by gravity $g$, and a characteristic value for the tension consequently is $\Theta = m_{s} g Z_{c}$. The solid is completely immersed so its effective mass per unit length reads $m_{t} = m_{s} + \rho \upi D_{eq}^2/4$, where an added mass coefficient of 1 is assumed, $\rho = 1000 \mbox{ kg.m}^{-3}$ being the fluid density. By defining a phase speed $c = \sqrt{m_{s} g Z_{c} / m_{t}} = \omega_{s} Z_{c}$, a characteristic length is derived as

\begin{equation}
	Z_{c} = \frac{m_{s}}{m_{t}} \frac{g}{\omega_{s}^2}.
	\label{Zc}
\end{equation}

\noindent{The dimensionless set of parameters consequently is}

\begin{equation}
	y = \frac{Y}{D_{eq}}, \quad	t = \omega_{s} T, \quad z = \frac{Z}{Z_{c}}.
	\label{dimensionless_eq}
\end{equation}

\noindent{With this scaling, the problem is characterized by one dimensionless coefficient, the string's reduced length $\ell = L / Z_{c}$, which depends on the flow speed and thus changes from one experiment to another.}

\subsection{Self-similar mode shapes}

For moderate values of $\Rey$, $\Rey \leq \Rey_{c}$ with $\Rey_{c} = 110$ for Chain 1, $\Rey_{c} = 160$ for Chain 2 and $\Rey_{c} = 250$ for Chain 3, vortex-induced standing waves are observed. Mode shapes are extracted for every experiment and are normalized using the $L^{2}$ norm. Figures \ref{Disp_brut} $(e)$ and $(f)$ show some similarity in the mode shapes of different experiments: the mode shape of the lower flow velocity case, figure \ref{Disp_brut} $(e)$, and the bottom part of the high velocity one ($Z < 0.10 \mbox{ m}$), figure \ref{Disp_brut} $(f)$, actually look alike. We therefore search here for a self-similar behaviour of the spatial distribution of the string's vibrations. Considering the mode shape of the experiment with the highest value of $\ell$ as a reference $\varphi_{ref} \left( z \right)$, the spanwise coordinate $z$ is rescaled for each experiment as $z/\kappa$, so as to minimize the $L^{2}$ norm of $\varphi \left( z/\kappa \right) - \varphi_{ref} \left( z \right)$:

\begin{equation}
	E^{2} \left( \kappa \right) = \int_{0} ^{\kappa \ell} \left[ \varphi \left( \frac{z}{\kappa} \right) - \varphi_{ref} \left( z \right) \right]^2 dz.
	\label{error}
\end{equation}

\begin{figure}
	\psfrag{phi}[tc][cc][1]{$\varphi$}
	\psfrag{KappaZ}[lc][cc][1][-90]{$\kappa z$}
	\psfrag{x00}[cc][cc][0.75]{0}
	\psfrag{xm4}[cc][cc][0.75]{-0.4}
	\psfrag{x04}[cc][cc][0.75]{0.4}
	\psfrag{y00}[cc][rc][0.75]{0}
	\psfrag{y40}[lc][cc][0.75]{40}
	\psfrag{y80}[lc][cc][0.75]{80}	
	\psfrag{y12}[cc][cc][0.75]{120}		
	\psfrag{a}[cc][cc][1]{$(a)$}		
	\psfrag{b}[cc][cc][1]{$(b)$}				
	\centerline{\includegraphics[width = 0.8\linewidth,height = 6cm]{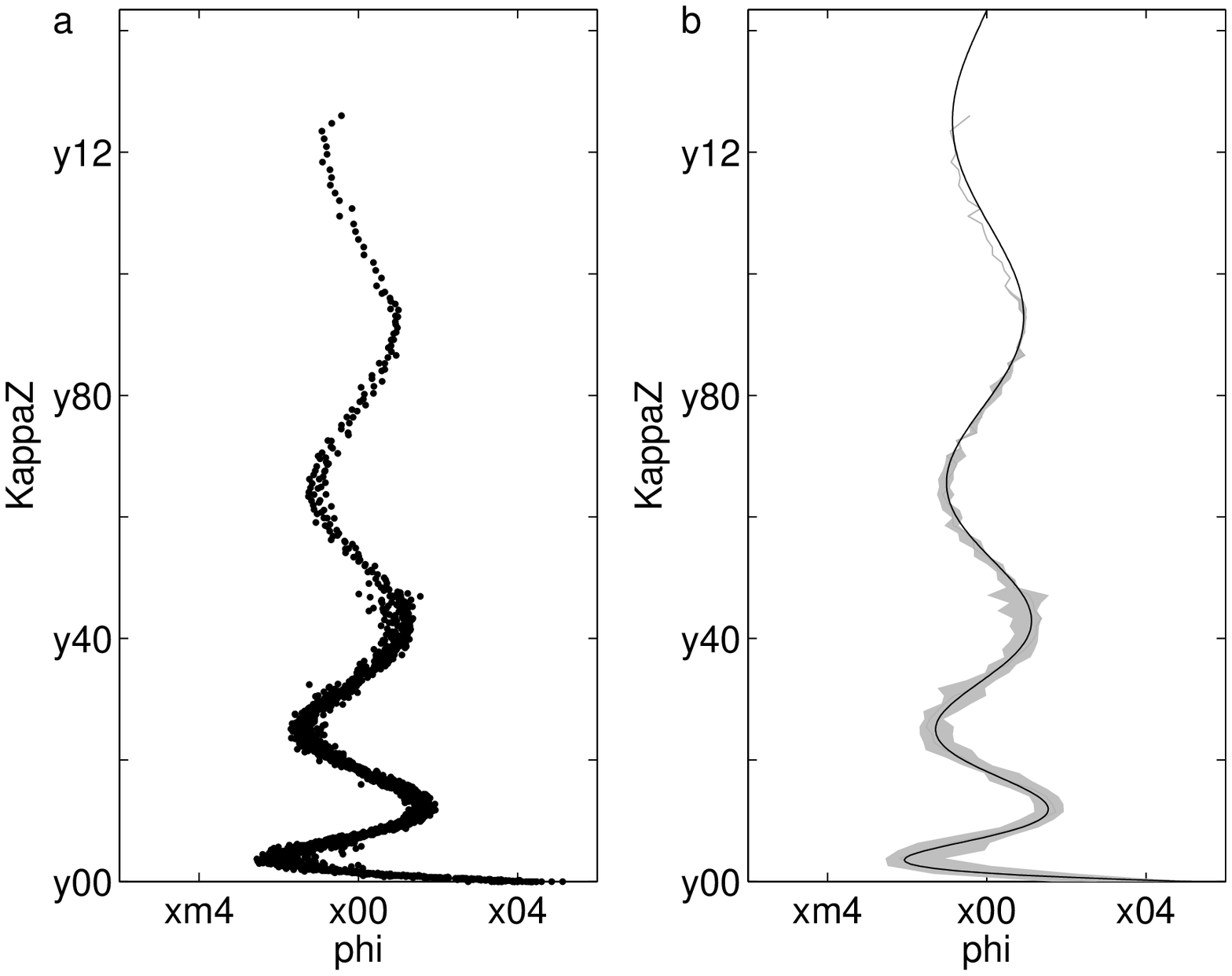}}		\vfill	
	\vspace{0.3cm}
	\psfrag{l}[tc][cc][1]{$\ell$}
	\psfrag{Kappa}[rc][cc][1][-90]{$\kappa$}
	\psfrag{x00}[cc][cc][0.75]{0}
	\psfrag{x40}[cc][cc][0.75]{40}
	\psfrag{x80}[cc][cc][0.75]{80}
	\psfrag{x12}[cc][cc][0.75]{120}	
	\psfrag{y05}[cc][cc][0.75]{0.5}
	\psfrag{y01}[cc][cc][0.75]{1}
	\psfrag{y15}[cc][cc][0.75]{1.5}	
	\psfrag{203mm}[cc][cc][0.65]{$D = 2.03$ mm }
	\psfrag{313mm}[cc][cc][0.65]{$D = 3.13$ mm}
	\psfrag{444mm}[cc][cc][0.65]{$D = 4.44$ mm}		
	\psfrag{c}[rc][lc][1]{$(c)$}		
	\psfrag{A}[cc][cc][0.75]{$A$}
	\psfrag{B}[cc][cc][0.75]{$B$}		  	
  \centerline{\includegraphics[width = 0.6\linewidth]{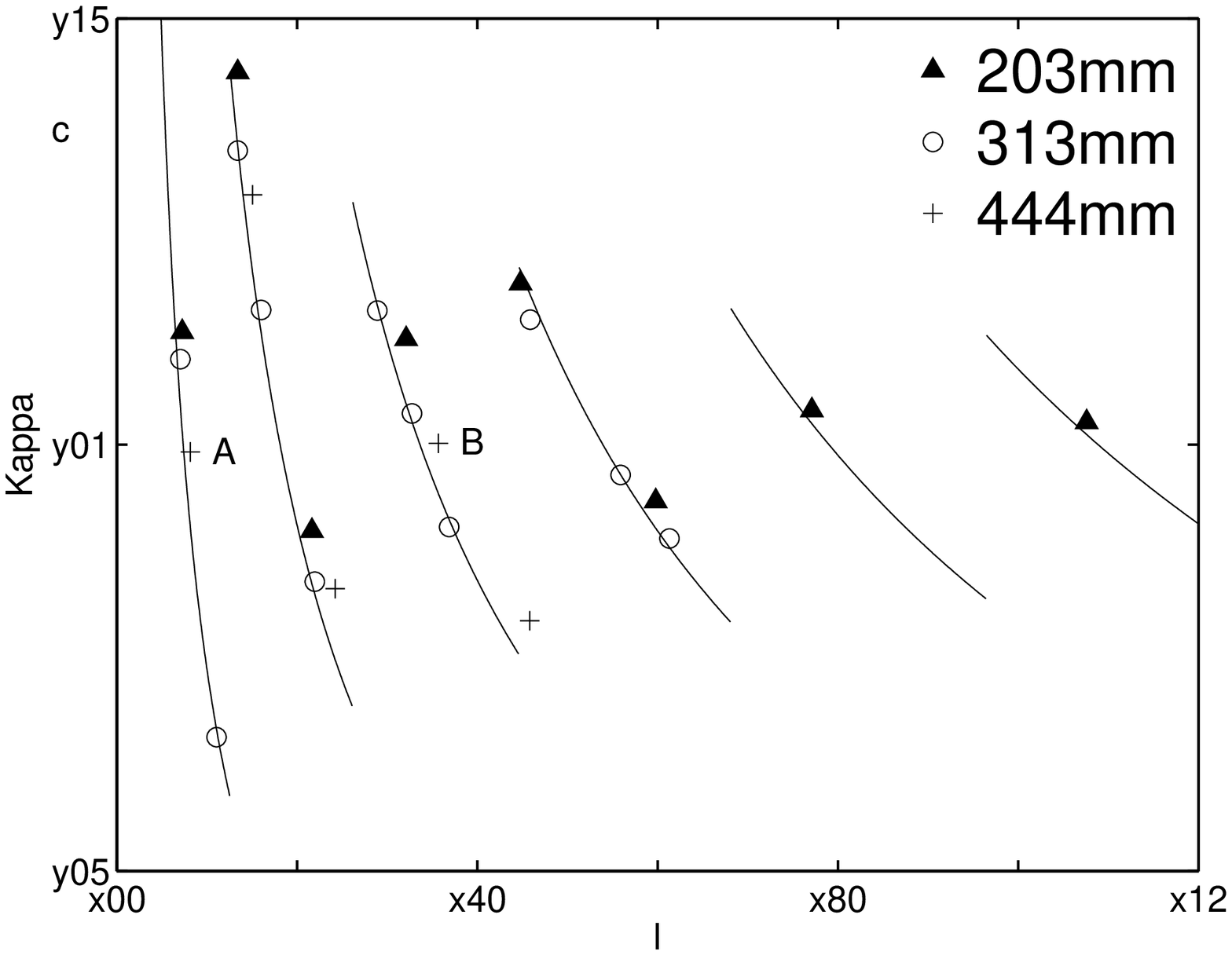}}
  \caption{Self-similarity of the mode shapes. $(a)$ Superposition of all the experimental mode shapes as a function of the rescaled spanwise coordinate. $(b)$ Comparison between the experimental mode shapes (grey) and the prediction of the linear model in section \ref{sec:section_model} (black continuous line). $(c)$ Evolution of the self-similarity coefficient $\kappa$ with the dimensionless length of the hanging string $\ell$, obtained from the experimental data (symbols) and using the linear model (continuous line)}
\label{self_similarity}
\end{figure}

\noindent{All the rescaled mode shapes are plotted as a function of $\kappa z$ in figure \ref{self_similarity} $(a)$. The collapse of all the mode shapes on the same function proves their self-similarity. As $\ell$ increases with $\Rey$, every experiment has a contribution to the lower region of the graph. There are consequently many more points at the bottom of the curve, which explains the higher scatter of the experimental data on this part of the plot.}

The evolution of the self-similarity coefficient $\kappa$ with $\ell$ is plotted on figure \ref{self_similarity} $(c)$. It strongly depends on the string's reduced length $\ell$, and exhibits many discontinuities. Each one of the continuous parts of this indented curve was observed to correspond to a different mode of vibration of the hanging string. For instance, the two cases shown in figure \ref{Disp_brut} $(a)$ and $(b)$ correspond to the points $A$ and $B$ in figure \ref{self_similarity} $(c)$, where the second and fourth structure modes are excited, respectively. The discontinuities in the evolution of the self-similarity coefficient $\kappa$ therefore seem to be associated with changes in the modes of vibrations. Further insight on this mode selection is provided in the next section using a wake-oscillator model.

\section{A linear wake-oscillator model} \label{sec:section_model}

Wake-oscillator models are reduced-order models \citep{Hart} that have been proven recently to render well the main features of the dynamics of a continuously flexible structure in VIV, in particular its frequency and mode shape \citep{Viol}. Moreover, the linear stability analysis of such a model has already shown a good ability to predict the dominant zone of each mode as well as its particular mode shape and frequency for tensioned cables in VIV \citep{Viol2}. This approach is adapted here to the case of a hanging string, where tension varies.

The dimensionless equation for the cross-flow displacement of the hanging string reads as

\begin{equation}
	\ddot{y} - \left( z y^{\prime} \right)^{\prime} = f_{fluid},
	\label{eq_solide}
\end{equation}

\noindent{where $f_{fluid}$ stands for the fluid forces on the solid apart from the added mass effect, which is already included in the definition of the effective mass per unit length $m_{t}$ \citep{Facc}. The fluid forcing is therefore proportional to a fluctuating lift coefficient, $f_{fluid} = Mq$ where $q = 2 C_{L} / C_{L0}$ and $M = C_{L0}/16 \mu \upi^{2} St^{2}$, $C_{L0}$ being the mean value of the fluctuating lift coefficient acting on the string and $\mu = m_{t} / \rho D_{eq}^{2}$ its mass ratio. The evolution of $q$ satisfies a harmonic oscillator equation forced by the string's dynamics,

\begin{equation}
	\ddot{q} + q = A \ddot{y}.
	\label{VdP}
\end{equation}

\noindent{Following Refs.~\cite{Viol2,Norb}, the values of the model parameters are taken as $A = 12$, $C_{L0} = 0.3$. Looking for solutions of the form $ \left[y,q \right] =  \Real{\left[ e^{i \omega t} \left( \varphi \left( z \right) , \psi \left( z \right) \right) \right]} $, it results from equation~\eqref{VdP} that $\psi = A \omega^{2} \varphi / \left( \omega^{2}-1 \right)$. Equation~\eqref{eq_solide} consequently becomes

\begin{equation}
\displaystyle \left( z \varphi ^{\prime} \right) ^{\prime}  + k^{2} \varphi = 0, \quad \\
  \label{Mode_shape_lin}
\end{equation}

\noindent{where the wavenumber $k$ and the frequency $\omega$ are linked through the dispersion relation}

\begin{equation}
\omega^{4} + \omega^{2} \left( AM - 1 - k^{2} \right) + k^{2}  = 0.
\label{dispersion}
\end{equation}

\noindent{The displacement of the string's lower end must remain finite, leading to the solution $\varphi \left( z \right) = J_{0} \left( 2 k \sqrt{z} \right)$ for equation~\eqref{Mode_shape_lin}, where $J_{0}$ is the zero-order Bessel function of the first kind \citep{Tria,Park}. The finite length of the string and the boundary condition at the fixed upper end restricts $k$ to discrete values, namely}

\begin{equation}
	\displaystyle k_{n} = \frac{\alpha_{n}}{ 2 \sqrt{\ell}},
	\label{wavenumber}
\end{equation}

\noindent{where $\alpha_{n}$ are the successive zeros of the Bessel function $J_{0}$.}

For given values of $n$ and $\ell$, the dispersion relation, equation~\eqref{dispersion}, provides the growth rate of an unstable mode as

\begin{equation}
	\displaystyle \sigma_{n} = \frac{1}{2} \sqrt{ - k_{n}^{2} + 2k_{n} + AM - 1 }.
	\label{growth_rate}
\end{equation}

\noindent{From equations~\eqref{wavenumber} and \eqref{growth_rate}, it can be found that mode $n$ is the most unstable mode for values of the dimensioness length $\ell$ in the interval}

\begin{equation}
  \displaystyle \frac{1}{16} \left( \alpha_{n-1} + \alpha_{n} \right)^{2} < \ell < \frac{1}{16} \left( \alpha_{n} + \alpha_{n+1} \right)^{2}.
	\label{interval}
\end{equation}
	
\noindent{The most unstable mode shape is $J_{0} \left( 2 k_{n} \sqrt{z} \right)$ with $k_{n} = \alpha_{n} / 2 \sqrt{\ell}$ over the interval defined by equation~\eqref{interval}. This form is consistent with the self-similarity of the mode shapes which has been observed experimentally in the previous section. The mode shapes are truncature of a unique function $J_{0} \left( 2 k_{n} \sqrt{z} \right)$ and the self-similarity coefficient is linked with the wavenumbers $k_{n}$ by the following relation

\begin{equation}
	\displaystyle \kappa = \left( \frac{k_{n}}{k_{ref}} \right)^{2},
	\label{kappa_k}
\end{equation} 
	
\noindent{where $k_{ref}$	is the most unstable wavenumber corresponding to $\varphi_{ref}$. The value of $k_{ref} = 1.017$ is derived according to the preceeding analysis, equation~\eqref{wavenumber}. The most unstable linear mode is shown in figure \ref{self_similarity} $(b)$, in excellent agreement with the mode shapes obtained experimentally. Furthermore, the values of the self-similarity coefficient $\kappa$ obtained analytically, equation~\eqref{kappa_k}, are also plotted in figure \ref{self_similarity} $(c)$. Again, the agreement of the predictions obtained using the linear wake-oscillator model with experiments is excellent. The stability analysis explains and predicts analytically each feature of the dependence of $\kappa$ with $\ell$, namely the origin and location of the discontinuities (change in the most linearly unstable mode) and the variations of $\kappa$.

The real part of the most unstable frequency is also derived in the same manner as the growth rate but is not shown here for brevity. Its evolution with $\ell$, centered around a value close to 1, exhibits some discontinuities because of the mode selection, similarly to the wavenumber $k$. The impact of the mode selection may explain part of the dispersion of the Strouhal number measurements in figure \ref{Strouhal_effectif}.}
	
\section{Conclusion}\label{sec:conclusion}

VIV of flexible slender bodies are characterized by the possibility of successive excitation of higher structural modes as the flow velocity is increased. In this paper, the dynamics of a hanging string in VIV is investigated experimentally to explore the effect of non-uniform tension. Over the range of $\Rey$ considered, standing vortex-induced waves were observed, with a concentration of the highest amplitudes in the regions of lowest tension. Excitations of up to the 7th mode of the string were observed for relatively low $\Rey$. The experiments prove the self-similarity of the spatial distribution of the vibrations and these dynamics were fully characterized. Furthermore, the variations of the self-similarity coefficient with the reduced length of the string was explained, using a linear stability analysis, as the result of the successive lock-ins of the different eigenmodes of the string. This standing wave structure breaks down at higher $\Rey$: travelling waves are indeed observed in the lower part of the chain that exhibits the largest amplitude of vibrations. This phenomenon, that is beyond the scope of the present paper, should be further studied to identify wether it is intimately linked to the system used in the present experiments or a fundamental property of hanging strings in high $\Rey$ flows.




\end{document}